\documentclass[
    ,final            
  ]
  {aipproc}

\layoutstyle{6x9}

\begin{document}

\title{Blind surveys for radio pulsars and transients}
\def\lapp{\ifmmode\stackrel{<}{_{\sim}}\else$\stackrel{<}{_{\sim}}$\fi}
\def\gapp{\ifmmode\stackrel{>}{_{\sim}}\else$\stackrel{>}{_{\sim}}$\fi}

\classification{}
\keywords      {}

\author{D.R.~Lorimer}{
  address={Physics Department, West Virginia University, Morgantown, WV 26506, USA}
}

\begin{abstract}
The main reasons for searching for pulsars are to: (i) get an accurate
census of the neutron star population and its origin and evolution; (ii)
connect neutron stars to other stellar populations in the Galaxy and 
globular clusters;
(iii) study Galactic astronomy (the interstellar medium and magnetic
field); (iv) find and study new interesting individual objects; (v) study pulsar
phenomenology; (vi) find pulsars to add to the sensitivity of pulsar
timing arrays.  This review focuses on blind (i.e. large area)
searches for radio pulsars.
I'll summarize the methods we use, some of the
challenges they present, look at some of the recent and current
efforts going on. I will also look at outreach of this area to groups
outside the traditional area of pulsar research, highlight the
discoveries of radio transients and look ahead to the future.
Pulsars found at other wavelengths will be reviewed elsewhere in this volume.
\end{abstract}

\maketitle

\section{Search methods}

The goal of pulsar searches is to find
a dispersed set of pulses buried in a two-dimensional data set of radio frequency
versus time. This is traditionally achieved by dividing the radio band
up into discrete frequency channels which are subsequently de-dispersed 
by appropriately delaying the frequency channels before summing over 
the whole bandwidth. Since in any given search we do not know the dispersion
measure DM\footnote{Defined, in the usual way, as the integrated column
density of free electrons along the line of sight to the pulsar, i.e.~$DM=\int_0^d n_e dl$.}
a-priori, we must create time series corresponding for a range of DMs.
To dig out periodic sources from the noise, we Fourier transform
each time series to reveal their signature as a series of harmonically
related components in the amplitude or power spectrum. We then search these
spectra for statistically significant features and repeat this process
for all trial DMs to obtain a list of pulsar candidates which are deemed
to be statistically significant. For each of these candidates, we return
to the original data set and fold it at the nominal DM and period and
examine the signal in time and radio frequency. An example set of diagnostic
plots showing a pulsar discovery
(in this case, from Scott Ransom's PRESTO software 
package\footnote{http://github.com/scottransom/presto}) can be seen in Fig.~1.

\begin{figure}
  \includegraphics[width=0.9\textwidth]{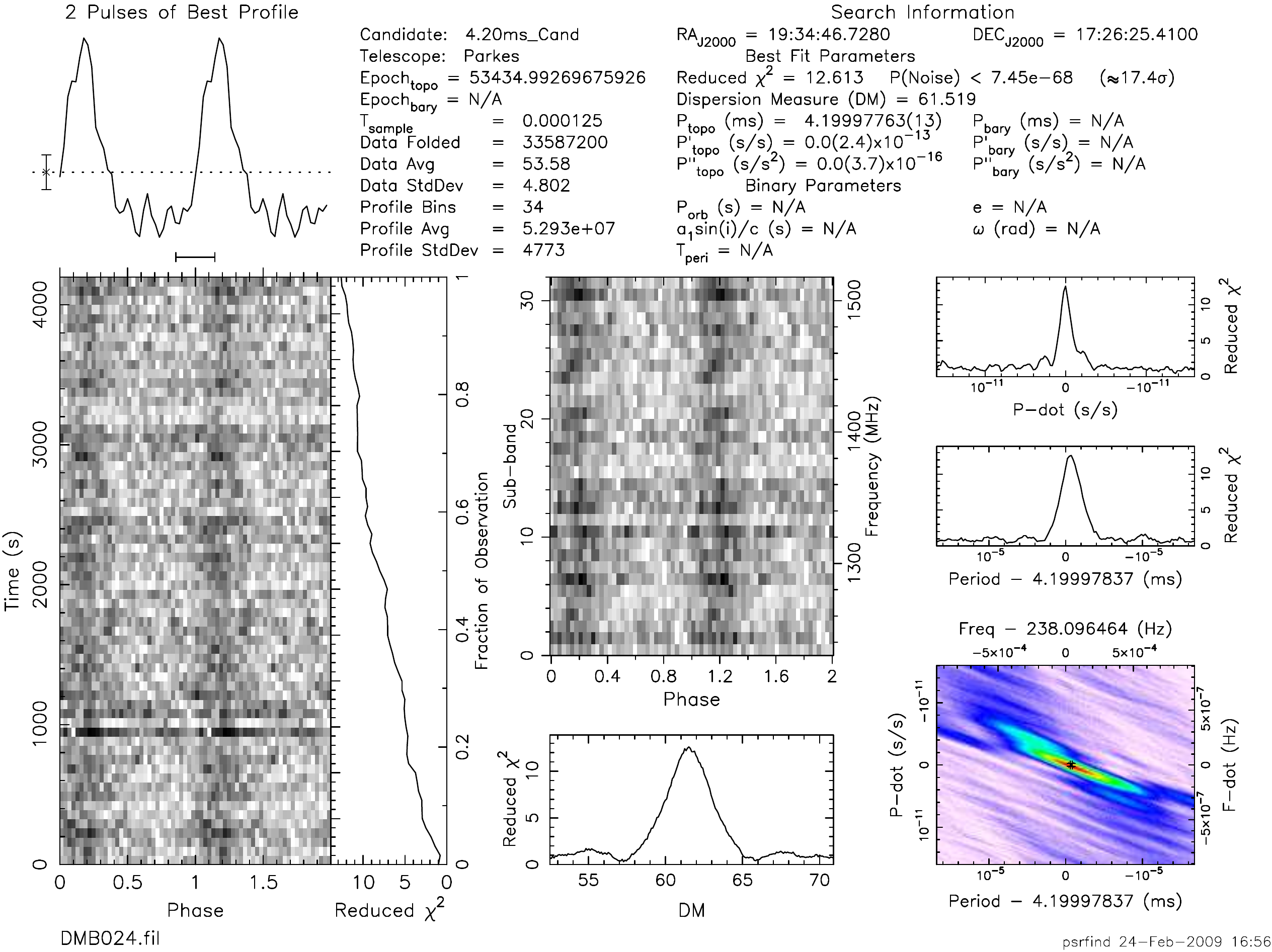}
  \caption{Example periodicity search output plot
showing data folded in time
(lower left) and radio frequency (upper center) as well as the integrated
pulse profile (upper left) and optimal DM search (lower center). 
The statistical significance of the signal in each of these diagrams
is measured in terms of the
reduced $\chi^2$ square value computed from the integrated pulse profiles.
A $\chi^2$ value close to unity would be found for a profile that is consistent
with Gaussian random noise. Shown here is
the discovery observation for the 4.2-ms pulsar J1935+1726
from a multibeam survey at
Parkes (Camilo et al.~2011).}
\end{figure}

\begin{figure}
  \includegraphics[width=0.9\textwidth]{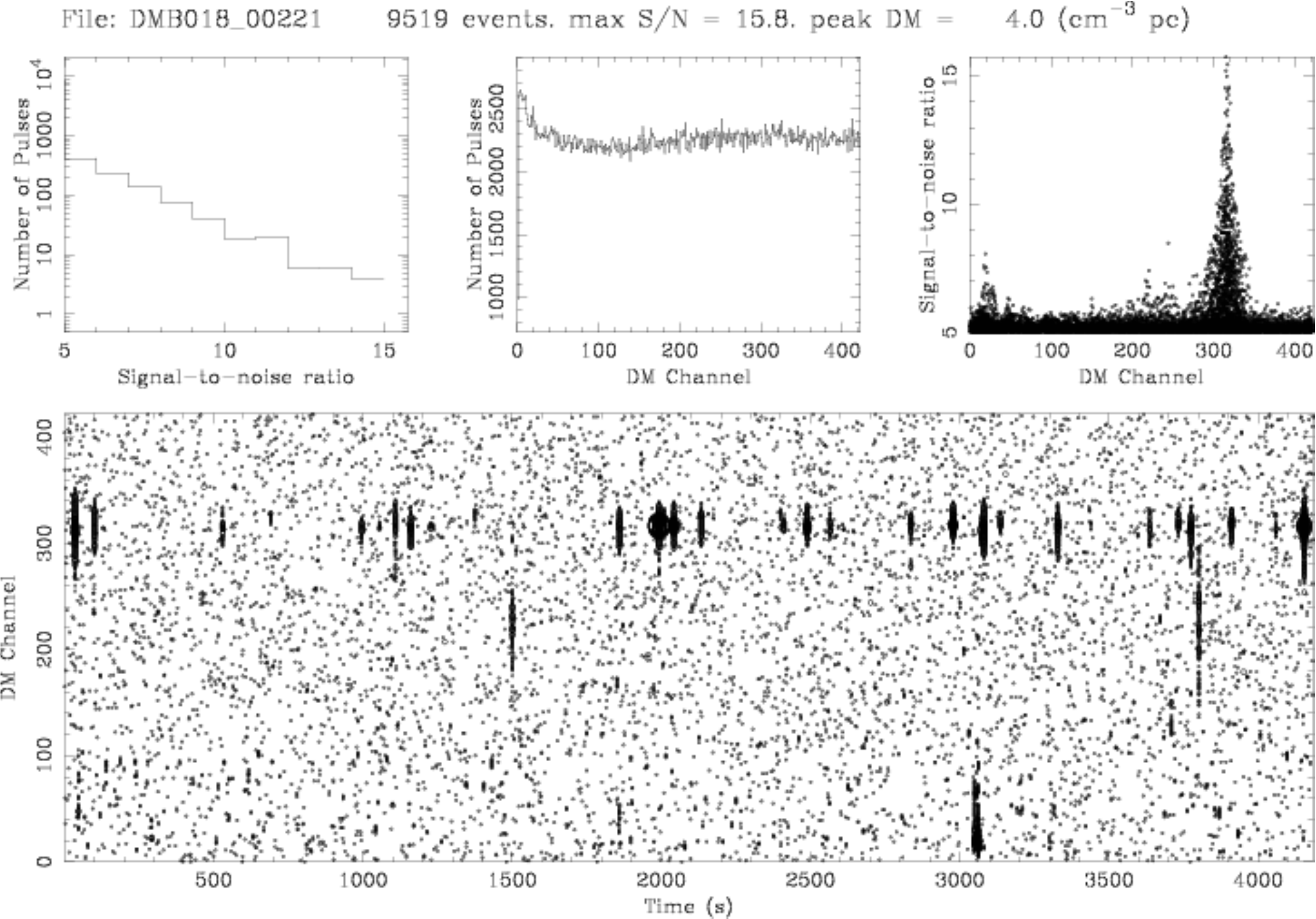}
  \caption{Example single pulse search plot, in this case from
Maura McLaughlin and Jim Cordes' matched filtering algorithm,
showing signal-to-noise of events as a function of trial DM
and time (lower plot), distribution of signal-to-noise
ratios (upper left), number versus trial DM (center) and
signal-to-noise versus trial DM (upper left). Shown here is
the discovery observation of the 1.9-s pulsar J1925+19
from a multibeam survey at
Parkes (Camilo et al.~2011).}
\end{figure}

The other method looks for individual pulses which would not appear
as significant features in the Fourier transform search. In this case,
one records individual events in each time series which significantly
exceed the overall rms noise level (e.g. 5$\sigma$ or above). The results
in this case is a set of diagnostic plots (in this case, from the SIGPROC
software package\footnote{http://sigproc.sourceforge.net})
shown in Fig.~2. From these,
it is fairly straightforward to see whether any significant signals
are present in the data. Further details on both search techniques
can be found in Lorimer \& Kramer (2005).

\section{Factors affecting search sensitivity}

The discussion here is necessarily brief, and highlights the key
limitations on a pulsar search. Further information can
be found in Lorimer \& Kramer (2005).

{\bf Scintillation} produces a diffraction pattern
on the observer's plane can result in significant flux density
modulation from one epoch to the next. This effect is particularly
significant when the scintillation bandwidth is greater than the 
observing bandwidth. It is particularly prevailant in low-frequency
($\lapp 1$~GHz) observations of low-DM ($\lapp 50$~cm$^{-3}$~pc)
pulsars and cannot be removed by instrumental means. Multiple 
observations of the same patch of sky are really the only way to
minimize this effect without changing the observing frequency
and/or bandwidth of the system. Multiple passes will also help mitigate
against pulsar intermittency.

{\bf Scattering} produced by multi-path
propagation as the pulses propagate through a turbulent interstellar medium
which results in rays arriving over a variety of path lengths and a 
classical one-sided exponential to pulse profiles. In extreme cases, the
signal-to-noise ratio (S/N) can be greatly reduced as the broadening
timescale approaches a pulse period. While this effect
cannot easily be removed via instrumental means, it can be minimized by
observing at higher frequencies since the exponential decay time
scales approximately as $f^{-4}$, where $f$ is the observing frequency. 
The trade-off here is that the pulsar
flux density spectra are generally very steep, scaling as $f^{-1.5}$.

{\bf Sky background} produced by free electrons radiating synchrotron
emission due to their motion in the
Galactic magnetic field is strongly position dependent. In particular,
it can add significantly to the overall system noise for observations
at low Galactic latitudes and towards the Galactic center. Fortunately
the spectrum of this radiation is also quite steep ($\sim f^{-1.6}$) so that
high-frequency ($f>1$~GHz) observations generally have only minimal 
contamination due to this effect.

{\bf Binary motion} during an integration is becoming increasingly
important to account for as we seek the most extreme binaries lurking
in the Galaxy. The shift in 
pulse phase during an observation of even just a few minutes
can be very significant
even for an orbital period of a few hours. For the most exotic binaries,
which might exist with orbital periods of a few 10s of minutes, only
by accounting for  orbital motion can one hope to detect such
a pulsar. These ``acceleration searches'' are carried out
in either the time or frequency domain and attempt to remove the 
effects of orbital motion as completely as possible. In the simplest
case, the pulsar's motion is approximated as a constant acceleration
during the orbit and this can result in substantial
gains in search sensitivity. For shorter orbital periods, where this
approximation breaks down, more sophisticated methods are required. One
such example is the demodulation approach being carried out by the
EINSTEIN@Home team on the PALFA data (Knipsel et al.~2010; see also
the contribution by Lazarus in these proceedings). This approach shows great
promise and should result in some very exciting discoveries over the 
next few years.

\section{Past and present surveys}

Many of the blind survey discoveries during the period 1998--2008 were
made by a variety of projects which took advantage of the Parkes 20~cm
multibeam receiver (Staveley-Smith et al.~1996). 
The 13 cooled receivers on the sky
allowed simultaneously a quantum leap in survey speed and sensitivity 
over most previous pulsar surveys. It has so far resulted in the discovery
of over 1000 pulsars. The various surveys responsible for this revolution
in discovery space are summarized in Table 1. 

\begin{table}
\begin{tabular}{llllr}
\hline
 \tablehead{1}{r}{b}{Surveyed region}&
 \tablehead{1}{r}{b}{Dwell time}     &
 \tablehead{1}{r}{b}{Area (deg$^2$)} &
 \tablehead{1}{r}{b}{Discoveries}    &
 \tablehead{1}{r}{b}{Primary reference}     \\
\hline
Galactic plane & 35~min & 1700 & $>700$ & Manchester et al.~(2001)\\
High latitudes & 265~s  & 3600 & 18     & Burgay et al.~(2006)\\
Intermediate latitudes & 265~s  & 2000 & 57     & Edwards et al.~(2001)\\
High latitudes  & 265~s  & 4150 & 26     & Jacoby et al.~(2009)\\
Northern Galactic plane   & 70~min & 40   & 18     & Camilo et al.~(2011)\\
Perseus spiral arm & 35~min & 500  & 13     & Burgay et al.~(2011)\\
Magellanic clouds & 2.3~hr & 270  & 14+3   & Manchester et al.~(2006)\\
\hline
\end{tabular}
\caption{Blind pulsar surveys carried out using the 20~cm Parkes multibeam
receiver and now the legendary one-bit filterbank spectrometers.
The three pulsars listed in the Magellanic clouds are new
discoveries in archival data from a recent analysis by the author
and summer student Z.~Bailey.}
\end{table}

Motivated by these discoveries, we are now enjoying an
era of high-sensitivity surveys carried out at a variety of telescopes.
These surveys are summarized in Table 2. The results from Parkes, where
the High Time Resolution Universe (HTRU) survey are now being undertaken
using a new generation of spectrometers on the Parkes multibeam system
are discussed by Keith and Bates in these proceedings. Further
contributions discussing the Green Bank drift-scan survey
(Boyles) and the Pulsar Arecibo L-Band Feed Array (PALFA) survey 
(Lazarus) are elsewhere in this volume. The Giant Metre Wave Radio telescope
(GMRT) is conducting a number of surveys at frequencies in the range
0.3--1.4~GHz. At Effelsberg, a northern complement
of the HTRU survey is being carried out (Barr, these proceedings).
The Green Bank Northern Celestial Cap (GBNCC) survey
is an ambitious project to survey the entire sky north
38 degrees in declination at 350~MHz using the GBT. A large number
of new discoveries are expected from these surveys over the next 
five years.

\begin{table}
\begin{tabular}{llllr}
\hline
 \tablehead{1}{r}{b}{Survey name}&
 \tablehead{1}{r}{b}{Telescope}     &
 \tablehead{1}{r}{b}{Frequency } &
 \tablehead{1}{r}{b}{Discoveries}    &
 \tablehead{1}{r}{b}{Primary reference}     \\
\hline
High Resolution Universe & Parkes & 1.4 GHz & $>65$ & Keith et al.~(2010) \\
Eight-grate Cygnus survey & Westerbork & 328 MHz & $>3$ & Janssen et al.~(2009) \\
Methanol Multibeam & Parkes & 6.8 GHz & 3 & Bates et al.~(2010)\\
Northern plane & GBT    & 350 MHz & $>33$   & Hessels et al.~(2009)\\
Drift-Scan     & GBT    & 350 MHz & $>27$ & Boyles (these proceedings)\\
Arecibo L-Band Feed Array & Arecibo & 1.4 GHz & $>55$ & Cordes et al.~(2006)\\
High Resolution Universe & Effelsberg & 1.4 GHz & 0 & Barr (these proceedings)\\
Northern Celestial Cap & GBT & 350 MHz & 0 & Ransom et al.~(2015)\\
Galactic plane & GMRT & 610~MHz & $>3$ & Joshi et al.~(2009)\\
\hline
\end{tabular}
\caption{The current generation of 
blind pulsar surveys carried out at a variety of observatories
and frequencies. Those surveys with zero discoveries listed
have only recently begun data collection. Those surveys with $>$ signs
listed are ongoing and further discoveries are expected soon.}
\end{table}

\section{Outreach efforts}

A number of groups have been recently been
engaging in involving high-school and undergraduate students in pulsar research.
In the Pulse@Parkes project (Hobbs et al.~2010), high-school students 
carry out regular timing observations of pulsars and learn about
their properties. The Arecibo Remote Control Center 
project\footnote{http://arcc.phys.utb.edu}
(ARCC; Miller et al.~2009) allows high-school and college students to 
search for pulsars and observe with the Arecibo and Green Bank telescopes.
An ``ARCC franchise'' has recently begun at the University of Wisconsin 
at Milwaukee (Xavier Siemens, private communication).

In West Virginia, a collaboration between West Virginia University
and the National Radio Astronomy Observatory has enabled a new
initiative known as the Pulsar Search 
Collaboratory\footnote{http://pulsarsearchcollaboratory.com} (PSC). The PSC,
which has been in existence since 2008, has involved high-school
students from WV and surrounding states in pulsar search data
from the GBT (Rosen et al. 2010).
A unique aspect of this project is that the data
are effectively ``owned'' by the students themselves and are not
being analyzed by other groups. The students can therefore aim to
make new discoveries and carry out the follow-up observations. Two
discoveries so far have been made of transient sources, the first of
which attracted significant interest from the White House with an
invite to President Obama's star 
party\footnote{http://www.nrao.edu/pr/other/obama} for discoverer Lucas 
Bolyard.

\begin{figure}
  \includegraphics[width=\textwidth]{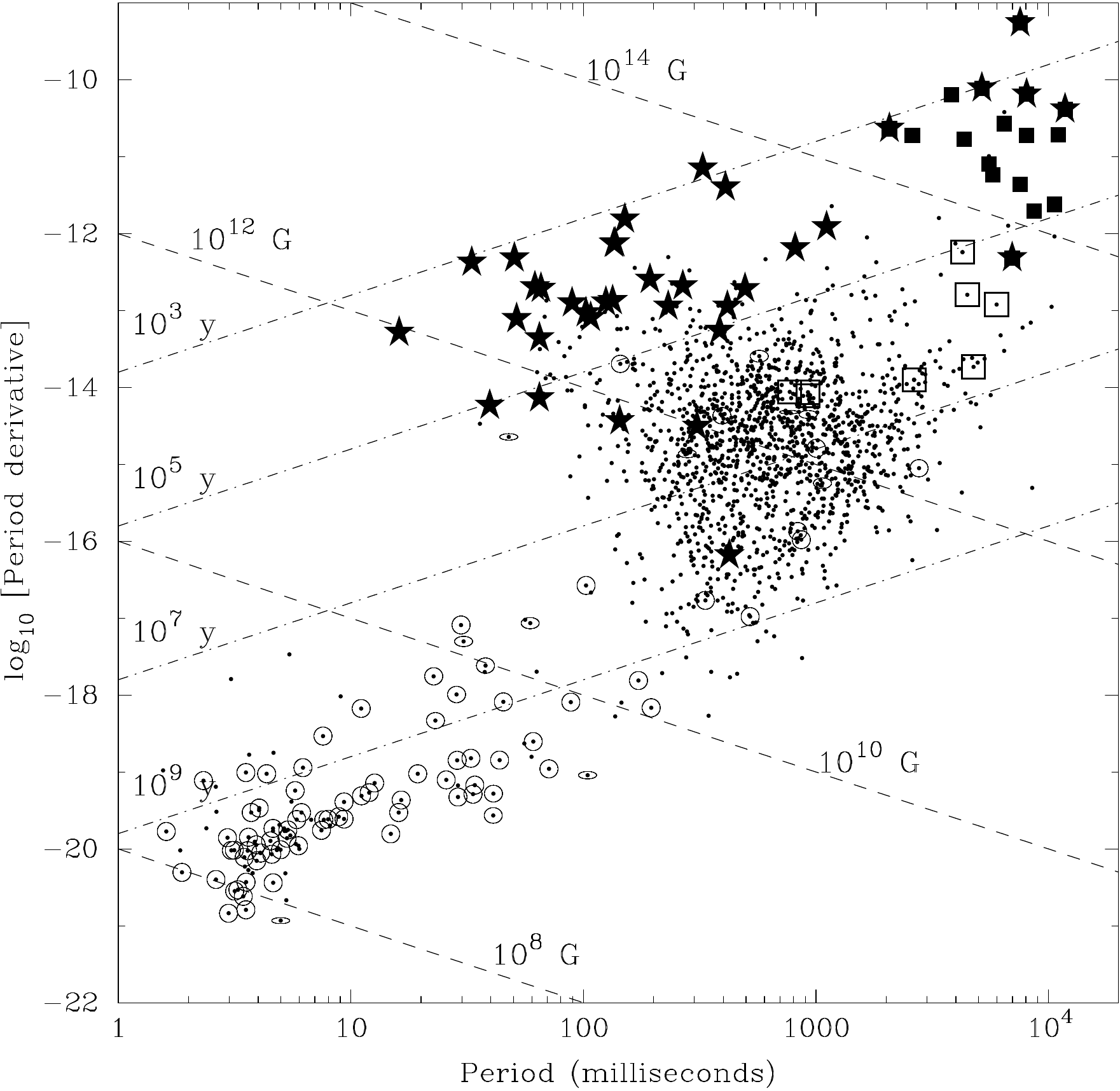}
  \caption{$P$--$\dot{P}$ diagram showing
normal pulsars as dots, binary pulsars as circles/ellipses,
magnetars as filled squares, RRATs as open squares and
supernova remnant associations as filled stars.
Lines of constant characteristic age $\tau \propto P/\dot{P}$,
magnetic field strength $B\propto \sqrt{P\dot{P}}$ are also shown.}
\end{figure}

\section{Transient sources}

Isolated dispersed radio pulses from neutron stars have been one of
the biggest new discoveries over the last decade or so.  Following
early work by Nice (1999) and Cordes \& McLaughlin (2001), it is now
commonplace in pulsar searches to include a single-pulse detection
module of the type described above. At Parkes, the discovery of the
rotating radio transients (RRATs) by McLaughlin et al. (2006) has
opened up a new window on the neutron star population. Of the 30 RRATs
currently known about a dozen now have measured spin-down rates and
their location on the $P-\dot{P}$ diagram shown in Fig.~4. While there
are many open questions as to how the RRATs fit into the pulsar
population as a whole (see the contributions by Miller and Kaspi in
these proceedings), one thing that is clear is that the RRATs manifest
themselves in a number of ways. Some are perhaps just ordinary pulsars
in which we see the tail end of the single-pulse amplitude
distribution (e.g. Weltevrede et al.~2006), while some are clearly
high $B$ and high $\dot{E}$ objects which show high-energy emission
and appear to more closely resemble magnetars than pulsars. In my
opinion, RRATs most likely represent a combination of neutron star
source populations rather than a single class. While they may be
observationally numerous as a whole, their birth rates need not add
significantly to the total neutron star formation rate if they have
evolved from other source classes which are already accounted for in
birth-rate estimates. For further discussion and views, see Keane \&
Kramer~(2008).

\section{Lessons learned}

As these surveys continue to produce exciting individual objects and
fill in our picture of the neutron star population, there are
lessons to be learned which will be useful to remember as
we move into the next era of searches with high-sensitivity facilities.

\begin{itemize}

\item{{\bf Always expect the unexpected.} The discovery of the RRATs is an
excellent example of how blind surveys continue to surprise us. In
general, casting as wide a net as possible when going on pulsar
fishing expeditions is clearly a good thing.}

\item
{{\bf Always archive your data!} The numerous re-analyses of the Parkes
Multibeam surveys using different software, new techniques and different
sets of eyes have really maximized the initial investment of telescope
time for these surveys. Although pulsar search data are typically
voluminous, it is extremely beneficial to invest in long-term archives
which procure the data for future processing.}

\item
{{\bf Involve groups outside of your field.} Our experience with high-school
students has noticeably changed perceptions about science within the groups
we have been working with and taps into a potentially vast resource of eyes
that are willing to look for pulsars in new and exciting ways.}

\item
{{\bf Experiment with machine-based candidate ranking schemes.} 
A number of groups are now actively applying the idea
of neural network based approaches to deal with the vast amount of output
that is now coming from pulsar searches (e.g. Eatough et al. 2010). As data
rates, sensitivity and sky coverage increase over the coming years, this
is set to become an even more important issue.}

\item
{{\bf Survey the sky multiple times with a given system.} As
the roles of pulsar intermittency and scintillation are becoming increasingly
recognized as limiting factors in pulsar search sensitivity, multiple
passes over each point in the sky can help mitigate these impacts.
Surveys carried out with comparable sensitivity in different epochs
(e.g.~spaced by a decade) have also shown that pulsar detectability
due to geodetic precession is an important factor (e.g.~Burgay et al.~2006).}

\end{itemize}

\section{Bold predictions for the future}

In conclusion, I would say that blind searches for pulsars have an
extremely promising future. Based on the current rate of progress of
existing surveys, it is reasonable to aim for a doubling of the
current population by 2015 (i.e.~a sample of around 4000 pulsars). If
surveys with more sensitive instruments such as FAST can be realized
by 2015, I predict that as many as 10,000 pulsars
could be found by 2020. Among these objects will surely be the
long-awaited pulsar black-hole binary.  As I tell the high-school
students each year, the pulsar-black hole binary system could be
discovered tomorrow by one of them! Another very tantalizing discovery
would be pulsars orbiting Sgr A*. Despite multiple searches to date
(Johnston et al.~2006; Deneva et al.~2009; Marquardt et al.~2010;
Bates et al.~2010), no radio
pulsar is currently known within 15 arcmin of the Galactic center. 
One of the great challenges is now to make the necessary sensitivity
improvements to detect a faint radio pulsar in this region.

\begin{theacknowledgments}

I thank the organizers for putting together a fantastic programme of
talks in a wonderful setting. My attendance at this meeting was enabled
by a Research Challenge Grant from the West Virginia EPSCoR program.
The ATNF pulsar catalogue (Manchester et al.~2005)
was used to make Fig.~3, along with
the McGill SGR/AXP Online 
Catalog\footnote{http://www.physics.mcgill.ca/$\sim$pulsar/magnetar/main.html}.

\end{theacknowledgments}


\begin{thebibliography}{10}

\bibitem{MMB}
S.~D. Bates et al.\ 2010, MNRAS, in press.

\bibitem{2006MNRAS.368..283B} 
M. Burgay et al.\ 2006, MNRAS, 368, 283.

\bibitem{pa} 
M. Burgay et al.\ 2011, in preparation.

\bibitem{2006ApJ...637..446C} 
J.~M. Cordes et al.\ 2006, ApJ, 637, 446.

\bibitem{2003ApJ...596.1142C} 
J.~M. Cordes, M.~A. McLaughlin, 2003, ApJ, 596, 1142.

\bibitem{2009ApJ...702L.177D} 
J.~S. Deneva, J.~M. Cordes, T.~J.~W. Lazio, 2009, ApJ, 702, L177.

\bibitem{2010MNRAS.407.2443E} 
R.~P. Eatough et al.\ 2010, MNRAS, 407, 2443.

\bibitem{2001MNRAS.326..358E} 
R.~T. Edwards, M. Bailes, W. van Straten, W. \& M.~C. Britton, 2001, 
MNRAS, 326, 358.

\bibitem{8gr8}
G.~H. Janssen et al.\ 2009, A\&A, 498, 223.

\bibitem{2009PASA...26..468H} 
G. Hobbs et al.\ 2009, PASA, 26, 468.

\bibitem{2009ApJ...699.2009J} 
B.~A. Jacoby, M. Bailes, S.~M. Ord, R.~T. Edwards, \& S.~R. Kulkarni, 
2009, ApJ, 699, 2009.

\bibitem{2006MNRAS.373L...6J} 
S. Johnston et al.\ 2006, MNRAS, 373, L6.

\bibitem{2009MNRAS.398..943J} 
B.~C. Joshi et al.\ 2009, MNRAS, 398, 943.

\bibitem{kk08}
E.~F. Keane \& M.~Kramer\ 2008, MNRAS, 391, 2009.

\bibitem{2010MNRAS.tmp.1356K} 
M.~J. Keith et al.\ 2010, MNRAS, 1356.

\bibitem{kac+10}
B. Knipsel et al., 2010, Science, 329, 1305.

\bibitem{2010ApJ...721L..33L} 
L. Levin et al.\ 2010, ApJ, 721, L33.

\bibitem{lcm10}
F. Camilo et al.\ 2011, in preparation.

\bibitem{lk05}
D.~R. Lorimer, M. Kramer, 2005, ``Handbook of Pulsar Astronomy'',
Cambridge University Press.

\bibitem{2001MNRAS.328...17M} 
R.~N. Manchester et al.\ 2001, MNRAS, 328, 17.

\bibitem{psrcat} 
R.~N. Manchester et al.\ 2005, AJ, 129, 1993.

\bibitem{2006ApJ...649..235M} 
R.~N. Manchester, G. Fan, A.~G. Lyne, V.~M. Kaspi \& F. Crawford, 2006, ApJ,
649, 235.

\bibitem{jp}
J.~P. Marquardt et al.\ 2010, ApJ, 715, 939.

\bibitem{2009AAS...21343104M} 
A.~F. Miller, F.~A. Jenet, A. Zermeno, K. Stovall, 2009, BAAS, 41, 264.

\bibitem{1999ApJ...513..927N} 
D.~J. Nice, 1999, ApJ, 513, 927 

\bibitem{GBNCC}
S.~M. Ransom et al.\ 2015, A\&A, in preparation.

\bibitem{PSC}
R. Rosen et al.\ 2010, AER, 9, 010106-1.

\bibitem{1996PASA...13..243S} 
L. Staveley-Smith et al.\ 1996, PASA, 13, 243.

\bibitem{2006A&A...459..597W} 
P. Weltevrede et al.\ 2006, A\&A, 459, 597.

\end{thebibliography}
\end{document}